\documentstyle[psfig]{mn}
\def\H0{{\it H}$_0$}
\def\Ms{{\it M}$_\odot$}
\def\Ls{{\it L}$_\odot$}
\def\q0{{\it q}$_0$}
\def\kmps{km~s$^{-1}$}

\def\ergps{erg~s$^{-1}$}
\def\kmpspMpc{km~s$^{-1}$~Mpc$^{-1}$}
\def\Ms{{\it M}$_\odot$}
\def\micron{$\mu$m}

\def\sec{$^{\prime\prime}$}
\def\nH{$N_{\rm H}$\thinspace} 
\def\psqcm{cm$^{-2}$}
\def\ergpspsqcm{erg~cm$^{-2}$~s$^{-1}$}

\def\Zs{$Z_{\odot}$}

\def\cps{ct\thinspace s$^{-1}$}

\def\Av{$A_{\rm V}$\/}
\def\pcubcm{cm$^{-3}$}
\def\ergcmps{erg\thinspace cm\thinspace s$^{-1}$}

\title[X-ray emission from NGC6240]
{ASCA spectroscopy of the luminous infrared galaxy NGC6240: 
X-ray emission from a starburst and a buried active nucleus}
\author[K. Iwasawa \& A. Comastri]
{\parbox[]{6.5in}{K. Iwasawa$^1$ and A. Comastri$^2$}\\
\\
$^1$ Institute of Astronomy, Madingley Road, Cambridge CB3 0HA\\
$^2$ Osservatorio Astronomico di Bologna, Via Zamboni 33 I-40126, Bologna, Italy\\}

\date{}

\begin{document}

\maketitle

\begin{abstract}
We present an X-ray spectroscopic study of the prototype far-infrared galaxy
NGC6240 from ASCA. The soft X-ray spectrum (below 2 keV)
shows clear signatures of thermal emission well described 
with a multi-temperature optically-thin plasma,
which probably originates in a powerful starburst. 
Strong hard X-ray emission is also detected with ASCA
and its spectrum above 3 keV is extremely flat with a prominent iron K
line complex, very similar to that seen in the Seyfert 2 galaxy NGC1068
but about an order of magnitude more luminous ($L_{\rm 3-10keV}\approx 1.4\times
10^{42}$\ergps).  
The hard X-ray spectrum indicates that only reflected X-rays 
of an active galactic nucleus (AGN) buried in a heavy obscuration 
(\nH$>2\times 10^{24}$\psqcm) are visible.
This is evidence for an AGN in NGC6240 emitting possibly at a quasar luminosity
($\sim 10^{45}$\ergps) and suggests its significant contribution
to the far-infrared luminosity.
\end{abstract}

\begin{keywords}
\end{keywords}

\section{introduction}

NGC6240 is one of the prototype far-infrared galaxies (FIRGs) that emit
most of their bolometric luminosities ($> 10^{11}$\Ls) in the far-IR
waveband (Soifer et al 1984).  The origin of the large IR luminosities
of this class of objects, comparable to QSOs, has been a big issue
since their discovery with IRAS. Either a starburst or an active nucleus
embedded in dust is the likely source of the energy output, but which
is the major component is still uncertain. X-ray spectoroscopy,
particularly with the sensitivity of ASCA in the hard X-ray band,
where the obscuration becomes optically thin, can be a powerful tool
to probe the energy source. The X-ray properties of several
objects belonging to the powerful FIRG class have been investigated 
with ASCA and the results
will appear elsewhere. We present here the
results on the brightest object, NGC6240, for which a detailed
spectral study is possible.

The characteristic properties of powerful FIRGs are found in NGC6240.
The 8--1000$\mu$m luminosity is $L_{\rm IR}\sim 6.6\times 10^{11}$\Ls
(Sanders et al 1989).
The distance of the galaxy is assumed to be 100 Mpc throughout 
this paper, calculated from the cosmological redshift of 0.0245 
using the Hubble constant of \H0 = 75 \kmpspMpc.
Two nuclei separated by 1.8\sec\ were found in this galaxy 
by optical imaging (Fried \& Schulz 1983)
and the distorted galaxy main body is elongated along approximately 
the N--S direction (PA$\sim 25^{\circ}$) with a prominent dust lane,
indicating that the system is undergoing a galaxy merger.
The large mass of molecular
gas, measured with CO line emission,
of $M_{\rm H_2}\sim 2\times 10^{10}$\Ms
(Sanders et al 1991; Solomon et al 1997) is 
confined within the projected size of $\sim 0.8$ kpc.

A deep CO absorption band at $2.3\mu$m and a powerful K-band continuum
($M_K = -24$) are consistent with the presence of 
a large number of red giants and supergiants.
The reasonably successful modelling of the near-IR emission 
in terms of starburst stellar population suggests that 
a starburst is a major power source (e.g., 
Ridgway, Wynn-Williams \& Becklin 1994; Shier, Rieke \& Rieke 1996).
Complex filamentary structures of a luminous H$\alpha$+[NII] nebula 
($1.7\times 10^{42}$\ergps; extinction-uncorrected) 
with a total size of 50 $\times$ 60 kpc, 
morphologically resembling the well known M82 filaments (McCarthy, Heckman
\& van Breugel 1987) but being much larger in size and 
$\sim 30$ times more luminous, were
imaged by Heckman et al (1987). The kinematics within the emission-line 
nebula were studied in detail by Heckman et al (1990) and
evidence for a high galactic wind velocity ($\sim 1000$ \kmps) has
been found. In fact, the optical emission-lines show LINER type excitation
(Fosbury \& Wall 1979; Keel 1990), 
which is in good agreement with shock heating by the galactic outflow.
A very strong H$_2$ 1--0 S(1) emission line 
at 2.12\micron ~(e.g., Lester, Harvey
\& Carr 1988; Ridgway et al 1994) that is spatially extended and
peaked between the two nuclei 
(van der Werf et al 1993) also appears to be shock-heated.

Although many observations point to a starburst as a major energy source
in this galaxy, the ionizing photons deficit problem for
Br$\gamma$ has been pointed out (DePoy et al 1986) and the large 
[FeII]$\lambda 1.64$\micron /Br$\gamma$ ratio
can be reconciled only when an Initial Mass Function (IMF) 
with a cut-off at $\sim 25$\Ms is assumed
(van der Werf et al 1993), if one tries to explain the whole IR emission 
(and ionizing photons) with a starburst alone. 
Although the nuclear radio source of NGC6240 can be interpreted 
as powered by a starburst, the ratio of 1.4 GHz radio power and 
far-IR luminosity is slightly larger than the mean value for spiral and 
starburst galaxies (Colbert, Wilson and Bland-Hawthorn 1994).
Note that 
the stellar velocity dispersion measured from the CO band, has the 
highest value so far measured 
($\sim 350$ \kmps, Doyon et al 1994; Shier et al 1996).
The inferred kinematic nuclear mass is too large to match the luminosity
(Doyon et al 1994).
Bland-Hawthorn, Wilson and Tully (1991) suggested the presence of 
a massive dark core through a 
study of kinematics of H$\alpha$ gas, although the core does not
lie at the radio core but at the secondary dynamical system which is
located at $\sim 10$\sec\ East of the centre of the galaxy.

The ASCA result on NGC6240 has been reported briefly by Kii et al (1997)
and Turner et al (1997). We present a detailed spectral analysis 
using ASCA and ROSAT data and discuss the properties of X-ray emission
from a starburst and an active nucleus in this galaxy.

\section{Observations and data reduction}

\subsection{The ASCA data}

NGC6240 was observed with ASCA (Tanaka, Inoue \& Holt 1994) on 1994 March 27.
The Solid state Imaging Spectrometer (the SIS; S0 and S1) was operated 
in Faint mode using two CCD chips. The SIS data are mainly used for
the spectral analysis presented in this paper because of its better 
sensitivity to soft X-rays and spectral resolution than the Gas Imaging
Spectrometer (the GIS; G2 and G3). The net exposure time for each detector is 
approximately $4\times 10^4$ s. Observed count rates for the S0 and G2 
detectors are $4.9\times 10^{-2}$\cps ~and $2.8\times 10^{-2}$\cps,
respectively. 
The observed fluxes in the 0.5--2 keV and 2--10 keV bands are 
$f_{\rm 0.5-2keV}\simeq 6.4\times 10^{-13}$\ergpspsqcm\ and
$f_{\rm 2-10keV}\simeq 1.9\times 10^{-12}$\ergpspsqcm.

Standard calibration and data reduction techniques were employed, using
FTOOLS provided by the ASCA Guest Observer Facility. The spectral analysis
was carried out using XSPEC.
The thermal emission model for optically-thin, collisional ionization
equilibrium plasma, 
MEKAL (Mewe, Kaastra \& Liedahl 1996), is used in the spectral fits,
with solar abundances taken from Feldman (1992).
The photoelectric 
absorption cross sections are taken from Morrison \& McCammon (1983).
Quoted errors to the best-fit spectral parameters are 90 per cent confidence
regions for one parameter of interest.


\subsection{The ROSAT PSPC data}

NGC 6240 was observed with the ROSAT X--ray telescope (Tr\"umper 1983)
with the Position Sensitive Propotional Counter (PSPC) 
in the focal plane (Pfeffermann et al. 1986)
on two separate occasions, September 1992 and February 1993, for $\sim$
5.2 ks and 3.1 ks respectively.

Radial profiles for the two images 
in the $\simeq$ 0.1--2.4 keV range have been 
convolved with the PSPC point spread function (PSF) according
to the source spectral properties and taking into account the background
level. The results reported in Fig.~1 for the longer exposure, 
clearly show that the X--ray emission
is extended in agreement with previous claims (Fricke and Papaderos 1996).
The extensions are clear in the full PSPC energy range as well
as in the restricted hard (0.5--2.0 keV) range, while at the softest 
energies ($<$ 0.4 keV) the source is extremely weak and no firm conclusions
can be drawn. Almost identical results have been obtained for the shorter
exposure.

Source spectra were extracted from circular regions
with radii of 3.5 arcmin.
Background spectra were taken either from annulii centered on the source
or from circular regions uncontaminated by nearby sources with extraction
radii as large as 10 arcmin. 
The large extraction cells ensure good statistics for the
modelling of the background spectrum and allow us to average over
the background small scale spatial fluctuations. 
Different background regions have been extracted 
and compared. In all the cases the background was stable without 
any appreciable variability within the statistical errors.

The photon event files were analysed using the EXSAS/MIDAS 
software (version 94NOV, Zimmermann et al. 1993) and the extracted 
0.1--2.2 keV spectra were analysed using version 9.0 of XSPEC 
with the appropriate detector response matrix (MPE No. 36).

Given that 
the source flux measured in the two separate observations is the 
same whithin 10 per cent, and the modelled spectra are also
consistent, the two photon event files have been merged.
The analysis procedure described above has been applied to the 
merged photon event list. The resulting exposure time is 8382 s and the
background subtracted source count rate is 0.070 $\pm$ 0.004 counts s$^{-1}$.

\begin{figure}
\centerline{\psfig{figure=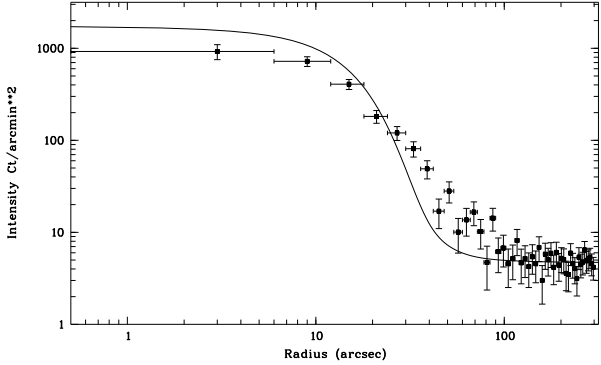,width=0.5\textwidth,angle=270}}
\caption{The ROSAT PSPC radial profile in the 0.5--2.0 keV band for the
5.2 ks exposure. The expected profile of a point source according to the
PSPC PSF and source spectrum is also reported.}
\end{figure}

\section{Spectral analysis}

The soft X-ray spectrum clearly shows characteristics of thermal emission
and above 3 keV, a hard tail emerges with a strong iron K
line feature on the continuum. Likely origins of the observed X-ray
emission are a starburst for the soft X-ray emission and scattered light 
from a hidden active nucleus for the hard X-ray emission, as shown below.

\subsection{The soft X-ray (0.5--2 keV) spectrum}

\subsubsection{The ROSAT PSPC data}

The source spectrum was rebinned in order to obtain a significant
signal-to-noise ratio ($>5$) for each bin and fitted either with a power law
or with a thermal plasma spectrum,   
plus absorption.
The derived spectral parameters are given in Table 1.

Both an absorbed steep power-law and a thermal model with 
low metal abundances plus excess absorption with respect to the 
Galactic value provide a good description of the observed spectrum.
Given the low statistics of the ROSAT spectrum, 
it is not possible to discriminate
between the two models. The power-law model is however ruled out from 
the high-resolution ASCA spectrum and the thermal emission model is 
favoured.


\begin{table}
\begin{center}
\caption{Spectral fits to the ROSAT PSPC data. The absorption column density
has been fixed at the Galactic value ($5.8\times 10^{20}$\psqcm).}
\begin{tabular}{cccc}
\multicolumn{4}{c}{Thermal emission model} \\
$kT$ & $Z$ & \nH & $\chi^2$/dof \\
keV & \Zs & $10^{20}$\psqcm & \\[5pt]
$1.27^{+0.69}_{-0.28}$ & $0.13^{+0.27}_{-0.13}$ & 5.8(fixed) & 12.2/12 \\
$0.83^{+0.55}_{-0.36}$ & $0.01^{+0.12}_{-0.01}$ & $14^{+23}_{-7}$ & 
7.8/11\\[10pt]
\multicolumn{3}{c}{Power-law model} & \\
$\Gamma$ & \nH & $\chi^2$/dof & \\
& $10^{20}$\psqcm & & \\[5pt]
$1.81^{+0.21}_{-0.22}$ & 5.8(fixed) & 20.3/13 & \\
$3.42^{+1.63}_{-1.08}$ & $27^{+29}_{-17}$ & 8.0/12 & \\
\end{tabular}
\end{center}
\end{table}


\subsubsection{The ASCA SIS data}


\begin{figure}
\centerline{\psfig{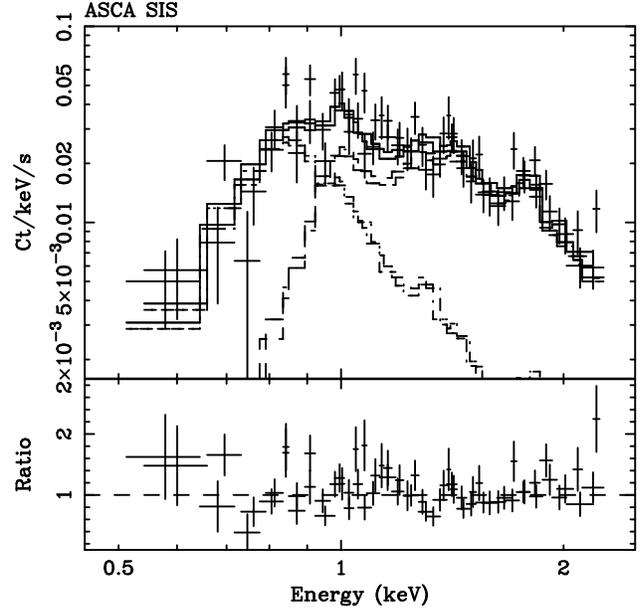}}
\caption{The ASCA SIS 0.5--2.3 keV spectrum of NGC6240. The two-temperature 
thermal emission model is fitted [Model (3) in Table 2].}
\end{figure}

The ASCA SIS spectrum shows a number of emission
lines from various ionized elements. Here we use the data in the 0.5--2.2 keV
energy range (Fig.~2), below the Au-edge due to the XRT, to cover 
emission lines from up to Si Ly$\alpha$ (2.0 keV).

We first fit the SIS data with a single temperature MEKAL model.
Whether the absorption is fixed at the Galactic value (\nH = 
$5.8\times 10^{20}$\psqcm) or left as a free parameter, 
fits are not satisfactory
leaving significant residuals suggestive of a more complex spectral
composition (see Table 2).
Introducing a two-temperature model, especially when absorption of the
higher temperature component is allowed to be a free parameter,
improves the fit to a reasonable quality. The metal abundance
is assumed to be identical in the two thermal components and 
the absorption for the low temperature component was fixed at the 
Galactic value [Model (3) in Table 2].
The two temperatures implied from the fit are $kT_1\approx 0.6$ keV and
$kT_2\approx 1.1$ keV, and the absorption for the high temperature 
component is $N_{\rm H,2}\simeq 1\times 10^{22}$\psqcm.

When absorption for the low temperature component is also left as
a free parameter,
the fit finds another set of $kT$ and \nH\ for the low 
temperature component [Model (4) in Table 2: $kT_1\simeq 0.23$ keV 
and $N_{\rm H,1}\simeq 4.6\times 10^{21}$
\psqcm, which is significantly larger than the Galactic column], 
leaving the parameters for the high temperature component 
to be almost the same, with a slightly worse $\chi^2$ value 
($\Delta\chi^2 = -2.8$). 

The main difference between the two models
in the relevant spectral range is found around 0.8--0.9 keV. 
Since the latter model underpredicts the emission in this band, 
significant line-like residuals remain.
Also no emission is expected in the latter model below 0.4 keV 
due to the large absorption, if
the covering factor of the absorption is unity. 

Unfortunately the source is barely visible in the ROSAT PSPC 
soft energy range (0.1--0.4 keV), so that it is very difficult 
to distinguish among the two models, which are both statistically 
consistent with the softest (0.1--0.4 keV) ROSAT PSPC counts..



\begin{table*}
\begin{center}
\caption{Spectral fits to the ASCA soft X-ray spectrum (0.5--2.3 keV)
of NGC6240. The metal abundance is assumed to be identical
in the two thermal components. $A$ is normalization of thermal emission model
($10^{-17}/(4\pi D^2)\int n^2 dV$) where $D$ is distance of the source in cm
and $n$ is electron density in cm$^{-3}$.}
\begin{tabular}{cccccc}
\multicolumn{6}{c}{Single temperature model} \\[5pt]
Model & $kT$ & $Z$ & \nH & $A$ & $\chi^2$/dof \\
 & keV & \Zs & $10^{21}$\psqcm & & \\[5pt]
(1) & $1.96_{-0.37}^{+0.63}$ & $0.09_{-0.06}^{+0.13}$ & 0.58 & 1.57 &
 184.9/125 \\
(2) & $1.54_{-0.38}^{+0.84}$ & $0.07_{-0.04}^{+0.09}$ & $1.1^{+1.0}$ & 1.92
& 183.4/124 \\
\end{tabular}
\end{center}
\begin{center}
\begin{tabular}{ccccccccc}
\multicolumn{9}{c}{Two-temperature model} \\
Model & $kT_1$ & $kT_2$ & $N_{\rm H,1}$ & $N_{\rm H,2}$ & $Z_{12}$ &
$A_1$ & $A_2$ & $\chi^2$/dof \\
& keV & keV & $10^{21}$\psqcm & $10^{21}$\psqcm & \Zs & & &\\[5pt]
(3) & $0.60_{-0.10}^{+0.07}$ & $1.09_{-0.19}^{+0.24}$ & 0.58 & 
$9.7^{+1.9}_{-0.22}$ & $0.27^{+0.23}_{-0.13}$ & 0.39 & 3.33 & 
150.9/122 \\
(4) & $0.24^{+0.06}_{-0.06}$ & $1.02^{+0.19}_{-0.23}$ & $4.5^{+2.4}$ & 
$10.3^{+3.1}_{-2.5}$ & $0.27^{+0.23}_{-0.13}$ & 5.1 & 3.8 & 153.7/121 \\
\end{tabular}
\end{center}
\end{table*}

\subsection{The hard X-ray spectrum and iron K feature}


\begin{figure}
\centerline{\psfig{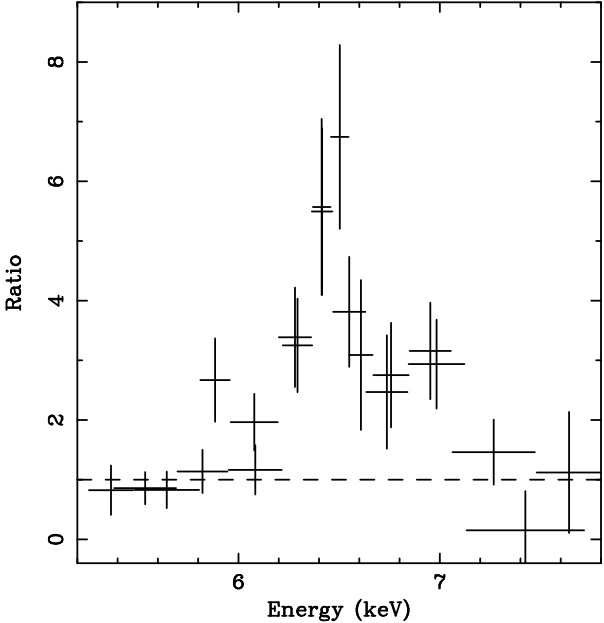}}
\caption{The iron K line profile in the SIS spectrum of NGC6240 displayed
in ratio to the power-law continuum. Two
peaks of emission are seen. The major line component is slightly broad
(see Fig. 4) perhaps due to a blend of multiple lines including a 
6.4 keV line. The energy scale is in the source rest--frame.}
\end{figure}


\begin{figure}
\centerline{\psfig{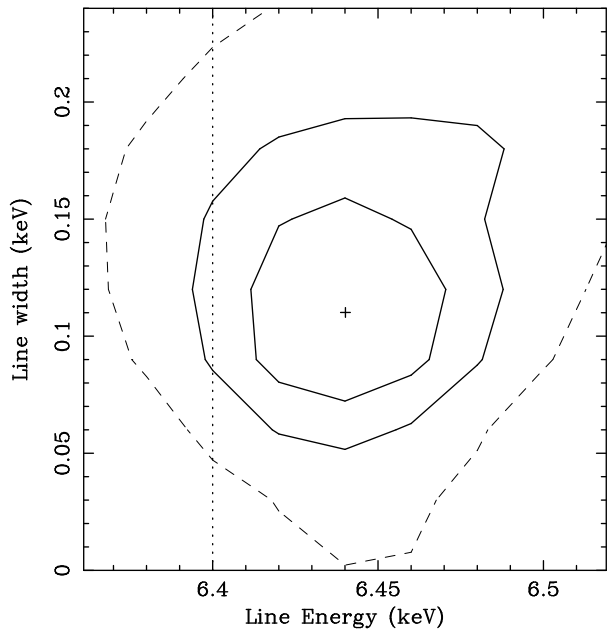}}
\caption{Confidence contours for line centroid energy and line width
of the major component (Line-1 in Table 2) 
of the iron K line complex of NGC6240.
The line energy is corrected for the redshift ($z=0.0245$) and the
line width is measured by a dispersion ($\sigma$) of a gaussian fit.
The spectral fits are performed for the SIS data.
The contour levels are 68, 90 and 99 per cent for two parameters of interest.
The energy of 6.4 keV expected from cold (less ionized than FeXVII) iron is
indicated by a dotted line.}
\end{figure}


\begin{table*}
\begin{center}
\caption{Line parameters of the iron K complex of NGC6240. A double-gaussian
model is fitted to the SIS data. $^{\ast}$ A single broad line is fitted,
which would be relevant when the line complex is observed with lower 
resolution spectrometers.}
\begin{tabular}{ccccc}
Fe K & $E$ & $\sigma$ & $I$ & $EW$ \\
& keV & keV & $10^{-5}$ph\thinspace s$^{-1}$cm$^{-2}$ & keV \\[5pt]
Line-1 & $6.44\pm 0.04$ & $0.11\pm 0.05$ & $2.9\pm 0.7$ & $1.58\pm 0.38$ \\
Line-2 & $6.87\pm 0.05$ & 0.01 & $1.3\pm 0.6$ & $0.75\pm 0.35$ \\[5pt]
Single$^{\ast}$ & $6.57\pm 0.07$ & $0.25\pm 0.10$ & $4.2\pm 0.9$ & 
$2.22\pm 0.48$ \\
\end{tabular}
\end{center}
\end{table*}

A strong iron K line feature around 6--7 keV is evident in the ASCA
spectrum (Fig. 3).  The underlying continuum above 4 keV is extremely
flat.  Fitting the 4--10 keV data excluding the iron K line band
(5.5--7 keV) with a power-law modified by Galactic absorption gives a
photon index of $\Gamma = 0.3^{+1.1}_{-0.8}$ from the SIS and $\Gamma
= 0.0^{+0.6}_{-0.4}$ from the GIS.  This very hard continuum is not
seen in any active nuclei or starburst galaxies except in the reflection
continua of Compton-thick Seyfert 2 galaxies such as the ones observed in
NGC1068 (Ueno et al 1994; Iwasawa et al 1997) and the Circinus galaxy
(Matt et al 1996).

The iron K line profile observed with the SIS is shown in Fig. 3.
There are at least two line components, well separated, and line parameters,
when two gaussians are fitted, are given in Table 3.

The stronger component centred at 6.44 keV is however broader than the
instrumental response. A gaussian fit provides
an instrinsic line width $\sigma = 110\pm 60$ eV.  Fig.~4 shows
a confidence contour plot for two parameters of the line, the rest
frame line energy and line width.  The best-fit line centroid energy is
marginally higher than 6.4 keV, the expected energy from
K$\alpha$ fluorescent line of cold iron (less ionized than FeXVII), at less
than 90 per cent confidence.  This perhaps means that the slightly
broadened line is a blend of the 6.4 keV line and another line at
a higher energy from iron of an intermediate ionization level.
Fitting with two narrow lines ($\sigma=10$ eV) one of which is at 6.4
keV, the other one is found at an energy of $6.54\pm 0.08$ keV
with a flux ratio to the 6.4 keV line of $\sim 0.6\pm 0.4$.

The identification of the 6.87 keV line is uncertain. The line
energy (see Table 3) is in between those of resonant lines expected
from FeXXV(6.70keV) and FeXXVI(6.97keV) in a highly photoionized medium, 
and indicate a blend of
the two. It could be FeXXVI redshifted by $\sim 4000$ \kmps, which is
similar to the feature (at $6.86\pm0.05$ keV) seen in the spectrum of 
NGC1068 (Iwasawa et al 1997). 
Higher signal-to-noise data (and spectral resolution)
are required to obtain secure identifications for the lines.

\subsection{The intermediate energy range: 2--5 keV}


\begin{figure}
\centerline{\psfig{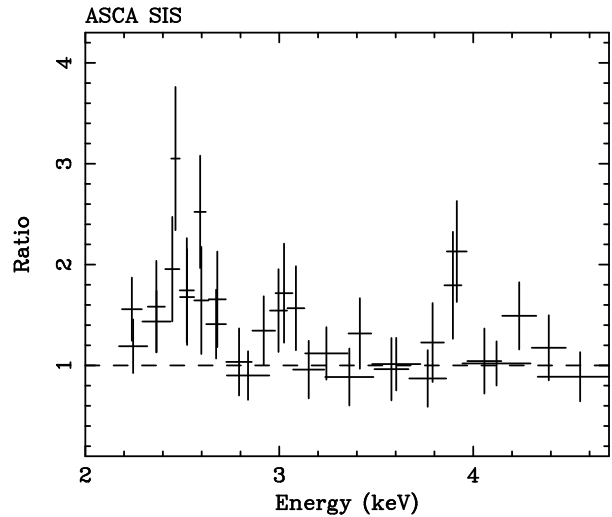}}
\caption{The emission line features in the intermediate energy range
of the ASCA SIS sepctrum of NGC6240. The ratio of the data to a power-law
($\Gamma = 1.47^{+0.49}_{-0.56}$ and Galactic absorption) is plotted. 
Significant emission line features are seen at 2.5 keV, 3 keV (ArXVII)
and 3.8 keV (CaXIX). The 2.5 keV line is broad due to a blend of Si and S 
lines. The energy scale is in the source rest--frame.}
\end{figure}


\begin{table*}
\begin{center}
\caption{Line parameters for intermediate energy emission-lines (Fig. 5). 
Line energies are corrected for the galaxy redshift ($z=0.0245$).
Equivalent widths are measured with respect to the underlying 
continuum modelled by a power-law (see text).}
\begin{tabular}{ccccc}
E & $\sigma$ & I & EW & ID \\
keV & keV & $10^{-6}$ph\thinspace s$^{-1}$cm$^{-2}$ & eV & \\[5pt]
$2.49\pm 0.06$ & $0.13\pm 0.07$ & $18^{+17}_{-10}$ & $358^{+336}_{-193}$ &
Si, S \\
$3.03\pm 0.06$ & 0.01 & $3.8^{+3.1}_{-2.9}$ & $100^{+81}_{-76}$ & ArXVII \\
$3.90\pm 0.05$ & 0.01 & $3.2^{+2.4}_{-2.3}$ & $125^{+93}_{-89}$ & CaXIX \\
\end{tabular}
\end{center}
\end{table*}

The spectral analysis above indicates that the soft X-ray spectrum below
2 keV is dominated by thermal emission due to a starburst while
the hard X-ray emission above 4 keV appears to be reflected AGN 
emission.
A transition between the two components occurs around 3 keV.
Several significant
emission lines are detected in the intermediate energy band 
(2.3--5 keV; Fig. 5), 
which can be used as a diagnostic of the origin of the X-ray emission. 

The 2.3--5 keV data are fitted with a power-law plus gaussians for 
emission lines. The photon index is $\Gamma = 1.47^{+0.49}_{-0.56}$ when
Galactic absorption is assumed. Line parameters and identifications 
for the three most prominent lines are given in Table 4. 
The 2.5 keV line is broad, indicating a blend of emission from
K$\beta$ lines of Si and K$\alpha$ of SXV(2.45 keV) and possibly 
SXVI(2.62 keV).
The origin of this line feature is probably the thermal emission, since
an extrapolation of the best-fit two-temperature model in Table 2 
accounts for it when the hard X-ray component is combined (see next section
and Fig. 6).
However, the extrapolation is far too low to explain the other two.
While the 3.9 keV line is consistent with emission from He-like calcium
(CaXIX), the 3.0 keV line is found at a slightly lower energy than expected
from He-like argon (ArXVII; 3.1 keV).

\subsection{Summary and interpretations of the whole band spectrum}



\begin{figure}
\centerline{\psfig{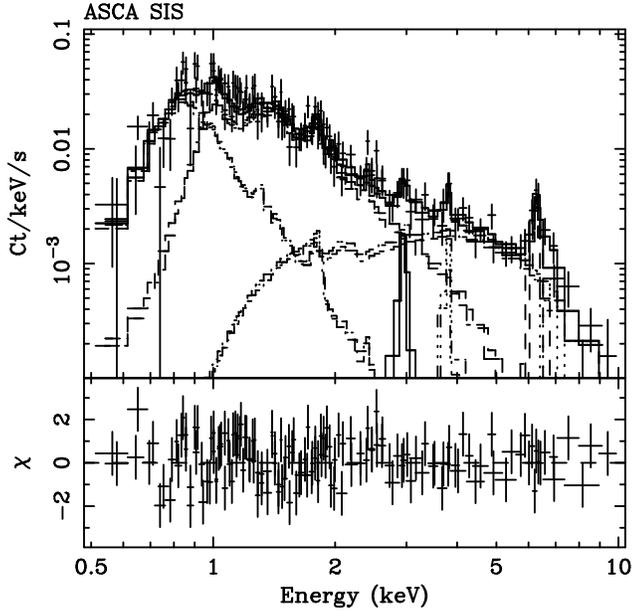}}
\caption{The 0.5--10 keV ASCA SIS spectrum of NGC6240 fitted with
a two-temperature MEKAL model plus an absorbed power-law with gaussians
for four significantly detected emission lines above 3 keV.}
\end{figure}

The soft X-ray spectrum can be described by thermal emission models
with two different temperature components (see Table 2). 
Significant absorption (\nH$\sim 1\times 10^{22}$\psqcm)
for the higher temperature component ($kT\sim 1$ keV) is required.
For the lower temperature one, extra absorption of order 
$\sim 4\times 10^{21}$\psqcm\ might be required if the gas temperature  
is $\sim 0.2$ keV, although $kT\sim 0.6$ keV gas with
no excess absorption gives a slightly better fit (see Section 3.1.2).
It is reasonable to assume that the higher temperature gas is 
confined in the inner region close to a starburst core 
where the obscuration is large. 
Assuming a standard gas-to-dust ratio the derived column density 
corresponds to a reddening $A_{\rm V}\simeq$ 4--5, which is
consistent with the optical estimate for the central core of the 
galaxy (\Av$= 4$) by Thronson et al (1990).

The thermal emission fades out around 3 keV and the spectrum becomes
flatter beyond that energy ($\Gamma_{\rm 2.3-5keV}\simeq 1.5$). 
Simple power-law fits suggest that the spectrum 
above 5 keV is even harder ($\Gamma_{\rm 4-10keV}\simeq 0.3$). 
The flat continuum around the Fe K line band and the large EW
of the 6.4 keV line are consistent with
reflection from optically thick, cold matter. However, 
the iron line components at energies higher than 6.4 keV 
can only be explained assuming reflection from 
highly ionized matter. 
It is difficult to estimate the ionization parameter of the 
material given the ambiguity in the identifications of the ionized Fe lines.
If the highest energy line is FeXXVI or a blend of FeXXV and FeXXVI with
similar intensities, the value of $\xi=L/nR^2$ would be 
a few times $10^3$\ergcmps (Kallman \& McCray 1982). 

The emission lines detected in the 3--4 keV range are probably
due to He-like calcium and argon, and no significant H-like lines for
those elements are observed. Since the underlying continuum is too flat
to match to any thermal emission model with such line properties 
(e.g., $kT\sim 2$ keV), 
a plausible origin for the emission in this energy range is reflected
light from ionized matter in a nuclear source possibly an active nucleus.
The detection of He-like lines of argon and calcium suggests that  
the value of $\xi$ should be $< 10^3$\ergcmps\ and more likely 
a few $10^2$\ergcmps.
Therefore the ionization state of this matter is perhaps lower than 
that for the ionized Fe line emitting matter.
The photoionized reflecting matter may have an ionization structure,
as observed in neutron star binary systems (e.g., Ebisawa et al 1996).
In this respect it is interesting to note that 
the reflection spectrum of NGC6240 is similar to the eclipse spectrum 
of Vela X-1 (Nagase et al 1994) in terms of the presence of the Fe K 
complex and He-like lines of lighter elements.

There is no evidence of a transmitted direct component, 
through obscuring matter, in the ASCA spectrum up to 10 keV.
This implies that the column density of the matter occulting 
the central X-ray source 
exceeds \nH\ $\sim 2\times 10^{24}$\psqcm, and the
detected hard X-ray emission is entirely due to reflection.
Here we model the ASCA 0.5--10 keV spectrum with a starburst plus an AGN
component. The two-temperature MEKAL model [Model (3) in Table 2] is 
again employed to describe a starburst emission. 
As a first approximation for the reflected AGN continuum spectrum 
we adopted a simple absorbed power-law. The detected emission lines 
($\S$ 3.2, 3.3) are also included and fitted with Gaussians. 

The best-fit values for the photon index and absorption column density are
$\Gamma = 0.3^{+0.6}_{-0.5}$ and \nH $=1.3^{+1.8}_{-1.3}\times 10^{22}$
\psqcm, respectively, and the parameters of the thermal emission are 
very similar to those in Table 2, except for the slightly higher 
metal abundance ($Z\sim 0.4$\Zs). The results are shown in Fig. 6.

The very hard spectrum of the AGN component can be explained by 
strong cold reflection, however some scattered emission
from an ionized medium should also be present as discussed above. 
In order to estimate the relative contributions of cold and ionized 
material, we also tried a model
in which  two reflectors (`cold' and `warm') are involved (Matt,
Brandt \& Fabian 1996). The spectral fit was performed in a similar way
described in Iwasawa et al (1997) for NGC1068. 
The photon index of the primary power-law source is assumed to be $\Gamma
= 1.8$.
The ratio of the cold and warm reflection is about 7/3 at 6 keV
and the 2--10 keV flux ratio is $\sim 1.5$, although uncertainties are quite 
large.
A higher energy continuum measurement with, for instance, 
the BeppoSAX/PDS could make a further constraint.

One could argue that the soft X-ray emission ($<2$ keV) is also due to
the scattered emission from ionized matter (e.g., Marshall et al 1993;
Netzer \& Turner 1997). However, 
this possibility seems to be unlikely due to the 
heavy obscuration over this dusty galaxy. Moreover the extension detected 
in the ROSAT PSPC image would imply an unreasonable high value for 
the size of the scattering region ($\sim$ 30 kpc). The starburst hypothesis
is thus favoured and could be tested by 
high spatial resolution imaging (i.e ROSAT HRI and/or AXAF).

\section{Discussion}

\subsection{X-ray evidence for a buried QSO}

The ASCA spectrum of NGC6240 is similar to that of NGC1068 in the
energy band above 3 keV.
The detection of a strong Fe K feature, particularly the 6.4 keV
line, is clear evidence for reprocessing of nonthermal emission in
cold gas. The ASCA spectrum strongly supports the presence of 
a powerful non-stellar
source hidden behind a thick obscuration with a column density
larger than $\sim 2\times 10^{24}$\psqcm.
The hard X-ray band may be the only window to see the AGN emission
in NGC6240. 
One could speculate that a powerful circumnuclear starburst
and the associated dust, would dominate the AGN emission at all wavelengths,
with the exception of hard X-rays and possibly the mid-infrared emission
via reradiation processes.
This luminous infrared galaxy thus appears to be an extreme case
in which intense star formation in the galaxy obscures its AGN emission.
A similar situation has also been found in the nearby starburst/Seyfert-2
galaxy NGC4945 (Iwasawa et al 1993; Done et al 1996).

The 3--10 keV luminosity of the reflection continuum is 
$1.4\times 10^{42}$\ergps,
an order of magnitude more luminous than NGC1068.  Since a likely
estimate of the intrinsic luminosity of the central source in NGC1068 is
$\sim 10^{44}$\ergps (e.g., Pier et al 1994), simple scaling predicts
a quasar luminosity $\sim 10^{45}$\ergps\ for
NGC6240, assuming a similar reflection geometry in the two nuclei. This
is comparable to the infrared luminosity ($2.5\times 10^{45}$\ergps).

The implied high column density of matter occulting the central nucleus
suggests that a large amount of cold gas surrounds the X-ray source.
The molecular gas inferred from radio interferometric observations, could be 
associated with this attenuating material.
Much of the CO emitting gas is concentrated in 
the central few hundred parsecs, and 
with their high density ($n\geq 10^4-10^5$\pcubcm; e.g., Solomon et al 1992)
it would be plausible to produce the high X-ray columns, 
if a number of clouds lie along our line of sight.

Perhaps the occultation of an X-ray source may occur on a much 
smaller scale. 
Recent radio interferometer observations (with water maser emission and
free-free continuum) revealed the presence of
an edge-on pc-scale disk in NGC1068
(Greenhill et al 1996; Gallimore et al 1997), which is likely to be associated
with the optically thick \nH\ $>10^{25}$\psqcm\ material obscuring the
central X-ray source (Koyama et al 1989; Matt et al 1997).

Despite the high detection rate of H$_2$O maser emission from
Seyfert 2 galaxies with column densities exceeding $10^{24}$\psqcm
(i.e. Compton-thick sources like NGC1068, NGC4945 and the Circinus galaxy),
no significant detection was reported for NGC6240 ($< 6.5$\Ls;
1$\sigma$ upper limit, Braatz et al 1996). 
A favourable geometry, e.g., edge-on, thin disk, may be required for
detecting strong water maser emission.

The dense molecular clouds are also responsible for the reflected 
X-rays and for the strong iron K flourescent line at 6.4 keV. 
Following Iwasawa et al (1997), the intrinsic 
$L^{\prime}_{\rm 3-10 keV}$ luminosity of a primary power-law source can 
be derived 
from the observed luminosity as :
$L^{\prime}_{\rm 3-10 keV} \approx f^{-1} \eta^{-1} 1.4\times 10^{42}$\ergps, 
where $\eta$=0.022 is the cold gas albedo 
and $f$ is the visible fraction of 
X-ray reflecting surface. Since the value of $f$ must be
much smaller than unity, it is evident that 
NGC6240 harbours an X-ray source with a quasar luminosity.

\subsection{X-ray emission from a starburst: a comparison with NGC1068}

NGC6240 is similar to NGC1068 in terms of the composite 
nature of starburst and reflected AGN emission. 
The soft X-ray emission in both galaxies is presumably attributed 
to a starburst.
If the soft X-ray emission ($<$ 2 keV) is entirely due to a starburst,
the 0.5--2 keV luminosity of NGC6240 ($\simeq 7\times 10^{41}$\ergps),
which is about 60 times more luminous than M82, is consistent with
a powerful starburst, as indicated by the luminous, large H$\alpha$ nebula.
It should also be noted that NGC6240 and NGC1068 have similar ratios
of the IR and soft X-ray (0.5--2 keV) luminosities 
[log (IR/SX)$\sim 3.6$].
However, there is a remarkable difference between their soft X-ray spectra.
NGC1068 has a huge soft X-ray excess and the spectrum below $\sim$ 2 keV 
is much steeper than that of NGC6240.

The ASCA soft X-ray spectrum of NGC1068 is fitted with a two-temperature
thermal model without absorption except at the Galactic value 
(Ueno et al 1994).
On the other hand, significant absorption of the order of 
$\sim 10^{22}$\psqcm\ is required, at least for the higher 
temperature thermal component, in NGC6240 ($\S$ 3.1.2). 
The presence of heavy
obscuration by dust in the starburst region in NGC6240 has been suggested by
optical and near-IR spectroscopy.
A similarly large obscuration has been found in the
nucleus of the nearby starburst galaxy M82 (e.g., Armus, Heckman \&
Miley 1990) 
and its soft X-ray spectrum 
observed with ASCA (Moran \& Lehnert 1997; Ptak et al 1997; Tsuru et al 1997)
is similar to that of NGC6240. In fact, the soft X-ray
spectrum of the central region of M82 is harder than that of the outer region,
suggesting either absorption in the nuclear region or the presence of 
low-temperature gas in the halo.

The emission-line properties in the optical and near-IR bands are
not at all similar to those of the narrow-line regions (NLRs) in typical
Seyfert galaxies or quasars, but consistent with shock heating by
starburst winds (Heckman et al 1990).
A NLR in NGC6240, which would be associated with a hidden AGN, is
invisible and probably obscured by a powerful starburst.

In constrast, the optical spectrum of NGC1068 is, of course, 
that of a Seyfert 2 galaxy and the reddening in the NLR is small. 
Also the ratio of near-mid IR and far-IR emission is larger in NGC1068 
than in NGC6240, indicating that hot dust is seen
in NGC1068 but not in NGC6240 (e.g., Pier \& Krolik 1992).
If obscuration is responsible for this difference, the required dust
extinction would be \Av\ $>5$ in NGC6240. 
A study of the ROSAT HRI image of NGC1068 (Wilson et al 1992) suggests that the
extended soft X-ray emitting region is almost absorption-free except for a part
shadowed by the circumnuclear molecular ring.

If the absorption of the thermal emission in NGC6240 is corrected,
the 0.5--2 keV flux increases by a factor of $\geq 4$.
This means that absorption by cold gas in the X-ray nebula could
attenuate the soft X-ray emission in NGC6240 otherwise an intrinsic
soft X-ray spectrum would have a large soft X-ray excess similar to
NGC1068. This correction would however lead the soft X-ray luminosity
of NGC6240 to be significantly larger than that predicted from conventional
starburst models (e.g., Chevalier \& Clegg 1985). Although previous
studies of far-infrared galaxies have shown that their soft X-ray luminosities
is typically four orders of magnitude below the IR luminoisties
(e.g., Griffiths \& Padovani 1990; Heckman et al 1990; 
David et al 1992), it is not clear that X-rays are produced in a similar
way also in the extreme objects like NGC6240. The high velocity 
($\sim 1000$\kmps) winds found in this galaxy and high density of X-ray
emittign medium could cause more efficient X-ray production. 

If the thermal X-ray emitting gas and cold absorbing gas 
are distributed uniformly over the nebula, the mass of the cold gas would be
unreasonably large ($\sim 0.9\times 10^{12}R_{10}^2N_{\rm H,21.5}$\Ms, 
where $R_{10}$ is a radius of the X-ray nebula in units of 10 kpc
and $N_{\rm H,21.5}$ is a column density in units of $3\times 10^{21}$\psqcm).
The X-ray emission could be emitted from a number of small
shock-heated clouds as suggested for M82 (Strickland, Ponman \&
Stevens 1997).

\subsection{Energy source powering the IR luminosity}

A starburst appears to power the total
emission from the optical to the near-IR band.
The optical emission-line spectrum is well explained with shock-heating
by Galactic-scale winds driven by a starburst.
The strong $2.2 \mu$m continuum and
the CO absorption band suggest that a starburst stellar population dominates 
the near-IR band (Ridgway et al 1994; Shier et al 1996).
The deep silicate absorption feature at $10\mu$m (\Av\ $\sim 50$) thus
arises in the molecular disk where a starburst is occuring.
The ISM in the starburst region would not be exposed to radiation
from a nonthermal central source if the source is surrounded by Compton-thick
matter, otherwise AGN characteristics would be observed in the 
near--mid-IR emission line spectrum.

In the mid-IR band, an AGN component possibly begins to emerge.
The ISO/SWS spectrum shows evidence for [OVI]$\lambda 25.9\mu$m, which 
is not commonly seen in starbursts but rather in AGNs (Egami et al 1997,
they also discuss a possibility of shock heating for the origin).
The far-IR spectrum is basically that of blackbody emission.
Although it is difficult to estimate the relative fraction of starburst 
and AGN components to the far-IR luminosity, given the X-ray evidence
for a powerful AGN, the AGN may power a significant fraction of the far-IR
emission.

In general, whether a quasar nucleus exists in other ultraluminous
IR galaxies is still unclear.
Apart from NGC6240, no positive X-ray signatures of a bright AGN  
have been found in
those objects which have no AGN-like optical emission-line properties
The ASCA spectrum of Arp220 shows thermal 
soft X-ray emission and very weak hard X-ray emission. 
If the ultraluminous IR galaxies contain a central source 
with quasar luminosity, it must be buried deep in dense gas clouds
with no emission from the central source escaping, otherwise 
some scattered light would be observed. It should be noted that none of the 
hyperluminous ($>10^{13}$\Ls) class objects have been detected with ASCA,
except IRAS 09104+4109 (Fabian et al 1994) in which the X-ray emission 
however originates in a cluster surrounding the galaxy (Fabian \& Crawford 
1995).
Perhaps central sources in most of the ultraluminous galaxies
have not evolved to powerful quasars. 
In fact, Mrk 273, an ultraluminous galaxy with Seyfert-2 type
optical properties, has an absorbed
X-ray source, but the absorption-corrected 2--10 keV
luminosity is only $2\times 10^{42}$\ergps, 3 orders of magnitude below 
the infrared luminosity (Kii et al 1996).
More details will
be discussed in another paper, together with several other prototype 
powerful far-IR galaxies.

\section{conclusions}

The main results of the ASCA and ROSAT observations of the luminous
infrared galaxy NGC6240 can be summarized as follows:

\begin{description}

\item[] (1) The ASCA spectrum at soft X-ray energies ($\sim$ 0.5--2.5 keV) is rather complex and can be fitted with two thermal components, likely to arise in a strong starburst, which is consistent with the detection of extended X-ray emission in the ROSAT PSPC image ($\sim$ 0.5--2.0 keV).

\item[] (2) The 3--10 keV spectrum is extremely hard and a strong iron K$\alpha$ complex feature is present. These results, very similar to those found for the Seyfert 2 galaxy NGC1068, indicate that NGC6240 contains a luminous active nucleus completely obscured by optically thick matter and 
only visible in reflected light. There has been little direct evidence for a powerful AGN in this object from previous observations at lower energies. 

\item[] (3) The inferred luminosity for the AGN component in the hard X-ray band suggests that a significant fraction of the far-infrared luminosity can be powered by the AGN. However, it is not clear whether the same scenario is applicable to ultraluminous infrared galaxies in general.

\end{description}

\section*{acknowledgements}

We thank all the members of the ASCA team who operate the satellite
and maintain the software and database. The referee is thanked for 
careful reading of the manuscript. The public ROSAT PSPC data were
retrieved from the ROSAT database at MPE.
KI thanks PPARC for support. Financial support from the Italian Space Agency 
under contract ARS-96-70 is acknowledged.

\end{document}